\documentclass[prl,twocolumn,showpacs,superscriptaddress]{revtex4}

\usepackage{color,psfig,graphics,psfrag,amsmath,amssymb,amsfonts,rotating}

\begin{document}

\title{Anderson Localization in Euclidean Random Matrices}

\author{S.~Ciliberti}

\affiliation{Departamento de F\'{\i}sica Te\'orica I, Universidad
Complutense de Madrid, Madrid 28040, Spain}

\affiliation{Instituto de Biocomputaci\'on y F\'{\i}sica de Sistemas
Complejos (BIFI). Universidad de Zaragoza, 50009 Zaragoza, Spain.}

\author{T.~S.~Grigera}

\affiliation{Instituto de Investigaciones Fisicoqu\'\i{}micas
Te\'oricas y Aplicadas (INIFTA), c.c. 16, suc. 4, 1900
La Plata, Argentina.}

\author{V.~Mart\'{\i}n-Mayor}

\affiliation {Departamento de F\'{\i}sica Te\'orica I, Universidad
Complutense de Madrid, Madrid 28040, Spain}

\affiliation{Instituto de Biocomputaci\'on y F\'{\i}sica de Sistemas
Complejos (BIFI). Universidad de Zaragoza, 50009 Zaragoza, Spain.}

\author{G.~Parisi}

\affiliation{Dipartimento di Fisica, Universit\`a di Roma ``La
Sapienza'', INFM unit\`a di Roma I, and Center for Statistical
Mechanics and Complexity (SMC), P.le A. Moro 2, I-00185 Roma, Italy}

\author{P.~Verrocchio}

\affiliation{Departamento de F\'{\i}sica Te\'orica I, Universidad
Complutense de Madrid, Madrid 28040, Spain}

\affiliation{Instituto de Biocomputaci\'on y F\'{\i}sica de Sistemas
Complejos (BIFI). Universidad de Zaragoza, 50009 Zaragoza, Spain.}


\begin{abstract}
We study spectra and localization properties of Euclidean random
matrices. The problem is approximately mapped onto that of a matrix
defined on a random graph. We introduce a powerful method to find the
density of states and the localization threshold.  We solve
numerically an exact equation for the probability distribution
function of the diagonal element of the the resolvent matrix, with a
population dynamics algorithm, and we show how this can be used to
find the localization threshold. An application of the method in the
context of the Instantaneous Normal Modes of a liquid system is given.
\end{abstract}

\pacs{PACS}

\maketitle


\paragraph{Introduction.} Some forty years ago Anderson~\cite{Anderson}
pointed out that disorder can turn a system expected to be a metal
from band-theory ({\sl i.e.\ } a gapless system) into an electric
insulator. This gave birth to the difficult problem of Anderson
localization~\cite{thou74, Abouchacra, Fyodorov, Lee}.  The physical
picture is roughly the following: In a pure system, states are
described by Bloch wavefunctions. Disorder divides the energy band in
localized (wavefunctions extending over a limited number of unit
cells) and extended regions (wavefunctions are plane-wave like,
involving an extensive number of lattice sites). The energy marking
this division is called {\em mobility edge} or localization
threshold. If the Fermi level lies in the localized region, transport
is strongly hampered. Given the large variety of systems that are
described to some degree of approximation by random
matrices~\cite{mehta}, it has become clear that Anderson localization
is significant beyond the physics of disordered metals. However, there
are very few situations~\cite{Abouchacra, CB} where it is possible to
estimate the position of the mobility edge in a controlled manner, and
often heuristic estimates are used (some authors identify it with the
Ioffe-Regel limit or with the limits of the spectrum obtained in some
kind of effective-medium approximation~\cite{BiMo}), or numerical
evaluations based on a variety of ideas (see {\sl e.g.}
refs.~\onlinecite{CarpenaFabianBeLa} or~\onlinecite{KraMac93}). In
this Letter we present a method to locate the mobility edge for
Euclidean Random Matrices~\cite{MPZ} (ERM), which we illustrate in a
liquid calculation of Instantaneous Normal Modes~\cite{INM} (INM).

The entries of an ERM are deterministic functions of (random) particle
positions.  Conservation laws (see below) are encoded in the
constraint that the sum of all elements in a row is null. ERMs appear
in the study of disordered $d$-wave superconductors~\cite{CHAMON},
disordered magnetic semiconductors~\cite{BRGS03} (very similar to a
spin-glass model~\cite{dean}), INM in liquids~\cite{cagi00,WuLo},
vibrations in glasses~\cite{noi}, the gelation transition in
polymers~\cite{BRODERIX} and vibrations in DNA~\cite{COCCO}.  Also,
theoretical studies have been carried out~\cite{MATHREF}.


\paragraph{ERMs in liquids.} Consider a translationally invariant system
with Hamiltonian $H[{\vec x}] = \sum_{i<j} v({\vec x}_i-{\vec x}_j)$,
for some pair potential $v({\vec x})$. The system has $N$ particles in
a box of volume $V$, the particle density being $\rho=N/V$.  We impose
a long-distance cut-off on $v({\vec x})$, as usual in numerical
simulations. In studies of short-time dynamics~\cite{INM} the harmonic
approximation is made around the oscillation centers ${\vec
x}^{c}$. We thus face the Hessian matrix $ M_{i\mu,j\nu}[{\vec x}^{c}]
\equiv - v_{\mu\nu}({\vec x}_i^{c}-{\vec x}_j^{c}) \ + \
\delta_{ij}\sum_{k=1}^N v_{\mu\nu}({\vec x}_i^{c}-{\vec x}_k^{c}) $,
where $v_{\mu\nu}({\vec x})$ is the second derivative matrix of
$v({\vec x})$. As it should, the sum of all elements in a matrix row
is zero.  Once the probability density function (pdf) of the
oscillation centers, $P[{\vec x}^{c}]$, is specified, the study of the
spectral properties of the Hessian is a problem on ERM
theory~\cite{MPZ}. In the context of supercooled liquids, people have
considered three types of $P[{\vec x}^{c}]$: equilibrium
configurations~\cite{INM} (the INM), minima~\cite{IS} and
saddle-points~\cite{SADDLES} of the potential energy surface. As in
all previous analytical studies, we restrict ourselves to INM.

\paragraph{The resolvent and its equation.} We need to consider the
resolvent matrix~\cite{thou74},
$R_{i\mu,j\nu}(z)=(z-M)^{-1}_{i\mu,j\nu}$ ($z$ is complex). To study
eigenvalues clustered around $\lambda$, we will set
$z=\lambda+\mathrm{i}\varepsilon$ for small $\varepsilon$. Our goal is
to find an equation for the pdf of the diagonal term of the resolvent
matrix~\footnote{Boldface symbols represent $3\times 3$ matrices
corresponding to particle displacements in 3-D.}, $\mathcal{P}[{\bf
R}_{ii}(z)]$, focusing its imaginary part. Consider the representation
of ${\bf R}_{ii}(z)$ in terms of the eigenvectors $|\alpha \rangle$ of
the matrix $M$:
\begin{equation}
  \text{Im}\, R_{jj}(\lambda+\text{i}\varepsilon) 
  = 
  \text{Im}\,
  \sum_\alpha \frac{|\langle j |\alpha\rangle|^2} {\lambda + \text{i}
  \varepsilon - \lambda_\alpha} \ .
  \label{diago}
\end{equation}
Previous calculations~\cite{cagi00,WuLo,BiMo} only addresed the {\em
mean value} of $\mathcal{P}[{\bf R}_{ii}(z)]$, which gives the density
of states (DOS) through $g(\lambda) = -\frac{1}{3 N\pi}\,
\text{Im}\,\overline{ \sum_i\text{Tr}\,{\bf
R}_{ii}(\lambda+\text{i}0^+)}$ (the overline indicates mean value).
Localization studies~\cite{thou74} require at least the variance of
$\mathcal{P}[{\bf R}_{ii}(z)]$, as can be seen from Eq.~\ref{diago}:
Consider a $\lambda$ in the localized region. The amplitude of
eigenvectors $|\alpha \rangle$ with $|\lambda_\alpha -\lambda|
\lesssim \varepsilon$ will be large for the main particle of the
eigenmode, and will decrease exponentially with the distance. Let
$n(\lambda)$ be the typical number of particles for which the
amplitude of an eigenmode is sizeable. The probability that in
Eq.~\ref{diago} particle $j$ will significantly participate in an
eigenvector of energy $\lambda_\alpha \in
(\lambda-\varepsilon,\lambda+\varepsilon)$ is of order $\varepsilon
g(\lambda) n(\lambda)$. In this case $\text{Im}\, R_{jj}$ is of the
order $1/(\varepsilon n(\lambda))$, while it is of order $\varepsilon$
with probability $1-\varepsilon g(\lambda) n(\lambda)$. Thus, the mean
value of $\text{Im}\, R_{jj}$ is finite for small $\varepsilon$ and
proportional to $g(\lambda)$, but its variance is
$\sigma^2(\lambda;\varepsilon) \simeq g(\lambda) / (\varepsilon
n(\lambda))$ and it diverges for vanishing $\varepsilon$.  A more
refined analysis~\cite{Abouchacra} shows that the pdf for $\text{Im}\,
R_{jj}$ decays as $(\text{Im}\, R_{jj})^{-\beta}$, $\beta\le 1.5$,
with an unknown cutoff function that prevents $\text{Im}\,
R_{jj}>1/\varepsilon$. On the other hand, in the extended region, all
the $|\langle j |\alpha\rangle|^2$ are $\mathcal{O}(1/N)$ and one can
replace the sum with an integral. It follows that the typical value of
$\text{Im}\, R_{jj}$ is a number of order one when $\varepsilon=0^+$.

To get an equation for $\mathcal{P}[{\bf R}_{ii}(z)]$, we have
considered the generalization of the Cizeau-Bouchaud recursion
relation~\cite{CB} to the case of translationally invariant
systems~\cite{long}. This equation relates ${\bf R}_{ii}(z)$ in a
system with N particles to the {\em full} resolvent of a system where
particle $i$ is not present anymore, but the positions of all other
$N-1$ particles are as in the original system. To use that equation we
need to make a rather drastic simplification regarding particle
correlations: following ref.~\onlinecite{cagi00}, we regard the matrix
elements $M_{i\mu,j\nu}[{\bf x}^{c}]$ as independent random variables,
$\mathsf{P}[{\bf M}]= \prod_{i < j} \mathsf{p}({\bf M}_{ij})$. Thus we
throw away most of the information contained in the Boltzmann weight
$P[{\bf x}^{c}]$, restricting ourselves to the pair correlation
function $g^{(2)}(r)$. Therefore, $\mathsf{p}({\bf M}_{ij})$ is the
pdf of the single matrix element,
\begin{eqnarray}
  \mathsf{p}({\bf M}_{ij}) &=&
  \!\left(1-\frac{\gamma}{N}\right)\!\delta({\bf M}_{ij})+\nonumber\\
&+&   \frac{\rho}{N}\int_{0}^{r^{\mathrm{cut}}} \!\!\!\!\!\!\! d{\vec r}\,
  g^{(2)}(r) \delta\left({\bf M}_{ij} +{\bf v}''(\vec r) \right) \,,
\label{DESACOPLADA}
\end{eqnarray}
where $\gamma=4\pi\rho\int_0^{r^{\mathrm{cut}}}\!\!\!\! dr'\,r'^2
g^{(2)}(r')$~\footnote{In the case considered here, $\gamma\approx
32$, thus the probability of finding disconnected clusters is
negligible~\cite{BrRo}.} is the average number of particles whose
distance from particle $i$ is less than the cut-off.  In other words,
we have considered the problem on a random graph~\cite{BrRo, long}:
each particle is the root of a Cayley tree whose fluctuating
connectivity is distributed with a Poisson law of mean value
$\gamma$. Under these hypothesis the central equation of this Letter
is obtained~\cite{long}:
\begin{eqnarray}
 \lefteqn{ \mathcal{P}[{\bf R}_{ii}(z)]  = 
  \int \!\! d\mathsf{P}[{\bf M}]\, d\mathcal{P} [{\bf R}_{jj}] \,
  \, \delta \left[ {\bf R}_{ii} -  \raisebox{0pt}[0pt][17pt]{} \right.   }
  && \label{CB3} \\
  && \left. \left(
    z + \sum_{j\neq i} {\bf M}_{i,j} -
    \sum_{j\neq i} {\bf M}_{ij} ({\bf R}^{-1}_{jj} + {\bf M}_{ij})^{-1}
	 {\bf M}_{ji}\right) ^{-1}
	 \right] .   \nonumber
\end{eqnarray}
Here the inverse-matrix symbols refer to the $3\times 3$ dimensional
space of the directions for particle displacements. Note also that all
the ${\bf R}^{-1}_{jj}$ are independently choosen from the
$\mathcal{P}[{\bf R}_{jj}(z)]$ distribution.

\begin{figure}
\centerline{ \includegraphics[width=\columnwidth]{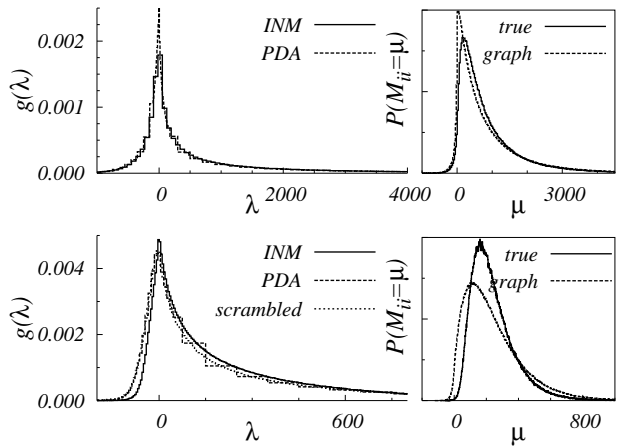} }
\caption{{\bf Top.} Left panel: DOS of a soft-sphere system at $\Gamma
= 0.6$ from the solution of Eq.~\ref{CB3} for $\varepsilon=1.0$
($N=20000$) and from the numerical diagonalization of the Hessian
matrix.  Right panel: distribution of the diagonal terms of the
Hessian matrix, from Eq.~\ref{DESACOPLADA} and from liquid simulations
at $\Gamma = 0.6$. {\bf Bottom:} As in the top part, for $\Gamma =
1.1$. The spectrum of the corresponding scrambled matrix (see text) is
also shown.}
\label{F-dos}
\end{figure}

\paragraph{Solving the equation.} We shall consider the
case of the INM of a monoatomic system with potential (in natural
units) $v(r)= (1/r)^{12}$, with a smooth cutoff at $r^{\mathrm{cut}} =
15\sqrt{3}/13$ (see Grigera {\sl et al.}\ in~\cite{SADDLES} for
details). As the only relevant thermodynamic parameter is $\Gamma =
\rho\,T^{-1/4}$, we shall take $\rho=1$. We have obtained the pair
correlation function by means of a Monte Carlo simulation of an
$N=2048$ system, for $\Gamma$ between 0.3 and 1.1, in the liquid phase
(cristalization happens at $\Gamma_c\approx 1.14$). We have also
obtained the INM spectrum numerically (diagonalizing 100 Hessian
matrices), in order to assess the quality of our approximations for
particle correlations. Once the $g^{(2)}(r')$ is known, one needs to
solve Eq.~(\ref{CB3}). We have done this numerically, by means of a
population dynamics algorithm (PDA).  Starting from a population of
$N$ elements with ${\bf R}_{ii}(z;t=0) = {\bf I}/z $, we iterate the
scheme
\begin{eqnarray}
  {\bf R}_{ii}(z;t+1)
  &=&
  {\bf F}\big({\bf R}_{jj}(z;t), {\bf M}_{ij}\big)
  \ ,\\
    {\bf F}\big({\bf R}_{jj}, {\bf M}_{ij}\big)
    &\equiv&
    \bigg[ 
      z{\bf I} + \sum_j {\bf M}_{ij} - \label{DINAMICA}
      \\
      [-3mm]
      && \sum_{j\neq i} {\bf M}_{ij}
      \big( {\bf R}_{jj}^{-1} + {\bf M}_{ij} \big)^{-1}
	  {\bf M}_{ji}
	  \bigg]^{-1}
    \ .\nonumber
\end{eqnarray}
To form the sums in (\ref{DINAMICA}), we divide the space around
particle $i$ up to the cutoff distance, in spherical shells of width
$\Delta r=r^{\mathrm{c}}/4096$. The probability of having a particle
in the shell is $4\pi\rho\int_r^{r+\Delta r}\!\!  dr'\,r'^2
g^{(2)}(r')$. Both the identity, $j$, of the particle interacting with
particle $i$, and the direction of the vector $\vec r\equiv \vec r_i
-\vec r_j$ are choosen randomly with uniform probability. A single
time-step consists on the sequential update of the full population.
We equilibrate for 100 time steps, which is verified to be enough by
checking the time evolution of the first moments of the
distribution. We then evaluate $\mathcal{P}[{\bf R}_{ii}(z)]$ by means
of (typically) 300 population-dynamics time steps. The population
averages of $\text{Im} R_{i\mu,i\mu}$ and $(\text{Im}
R_{i\mu,i\mu})^2$ are calculated, then time-averaged. From them we
compute the DOS, $g(\lambda)=-\overline{\text{Im} R_{i\mu,i\mu}}/\pi$,
and $\overline{\text{Im} R_{i\mu,i\mu}^2}\,$. An important test is to
compare the DOS obtained from Eq.~\ref{CB3} with the DOS obtained by
numerical diagonalization of the scrambled matrix $\tilde {\bf M}$,
built by picking its off-diagonal elements randomly and without
repetition from the off-diagonal elements of the true Hessian, and
then imposing the constraints of symmetry and translational
invariance.  This kills three-body and higher correlations, making
Eq.~\ref{CB3} exact. In Fig.~\ref{F-dos} we show the DOS at
$\Gamma=0.6$ and 1.1, computed both from Eq.~\ref{CB3} and from the
numerical diagonalization of the Hessian. At $\Gamma=0.6$ good
agreement is found. At $\Gamma=1.1$ we also show the DOS of the
scrambled Hessian, which agrees with Eq.~\ref{CB3}.  One can
understand the mild disagreement between Eq.~\ref{CB3} and the INM
spectrum at $\Gamma=1.1$ by comparing the pdf of the diagonal term of
the Hessian matrix as computed from Eq.~\ref{DESACOPLADA} and from the
actual liquid configurations (Fig.~\ref{F-dos}). This pdf coincides
with the DOS at the leading order in perturbation theory (the diagonal
term, ${\bf M}_{ii}$, is much larger than the typical off-diagonal
term ${\bf M}_{ij}$), so we understand why at $\Gamma=1.1$,
Eq.~\ref{CB3} overestimates the weight of the imaginary frequencies.

\begin{figure}
\centerline{ \includegraphics[angle=-90,width=\columnwidth]{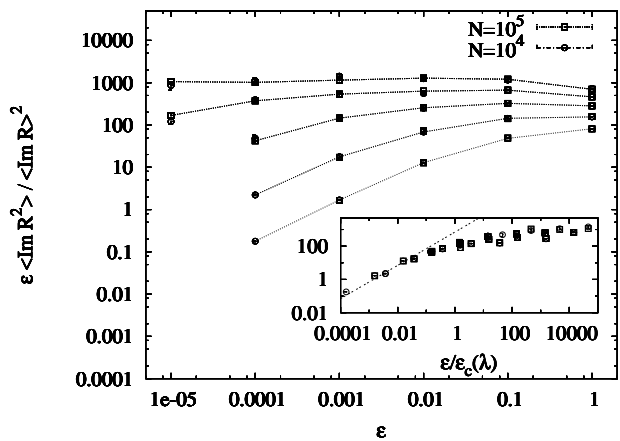}  }
\caption{$l(\lambda;\varepsilon)$ {\sl vs.\/} $\varepsilon$ near the
lower mobility edge at $\Gamma=1.1$ for (from top to bottom)
$\lambda=-140$, $-130$, $-120$, $-110$, and $-100$. Inset: Scaling
plot $l(\lambda;\varepsilon)$
vs. $\varepsilon/\varepsilon_c(\lambda)$ for $\lambda\ge -130$.}
  \label{F-scal11}
\end{figure}


\paragraph{Localization.} In principle, one can establish the
localization threshold from the behavior of the variance of the
imaginary part of $R$. Consider $l(\lambda;\varepsilon) \equiv
\varepsilon \langle \text{Im}R^2\rangle / \langle \text{Im}
R\rangle^2$. From the remarks above it follows that in the localized
region and for small $\varepsilon$, $l(\lambda;\varepsilon)$ tends to
a constant, while in the extended region the variance is finite and
$l(\lambda;\varepsilon)$ is
$\mathcal{O}(\varepsilon)$. Fig.~\ref{F-scal11} shows
$l(\lambda;\varepsilon)$ around the lower mobility edge. For the
higher values of $\lambda$, the extended, $\mathcal{O}(\varepsilon)$
regime is reached, but as the threshold is approached, the regime sets
in for lower and lower values of $\varepsilon$, making the threshold
difficult to estimate. To overcome this limitation, we have tried the
phenomenological scaling
$l(\lambda,\varepsilon)=f(\varepsilon/\varepsilon_c(\lambda))$, with
$f(x) \sim x$ for $x\sim 0$ and $\varepsilon_c(\lambda)\to0$ for
$\lambda\to\lambda_\text{th}$ (Fig.~\ref{F-scal11}). The
$\varepsilon_c(\lambda)$ (that we obtained only for $\lambda\ge -130$)
can be fitted to a power law $c(\lambda-\lambda)^\gamma$, with
$\gamma=21$ and $\lambda_\text{th}=-171.6$. This huge exponent is
perhaps an indication of an essential singularity at the mobility edge
(cf.\ the analitical results of ref.~\onlinecite{Fyodorov}).

However, the fit is not very reliable, since the it does not reach
close enough to the critical point (moving nearer requires lowering
$\varepsilon$, but this is computationally expensive since the
statistics must grow as $1/\sqrt{\varepsilon}$ when the variance is
growing as $1/\varepsilon$).

\begin{figure}
\centerline{\includegraphics[angle=-90,width=\columnwidth]{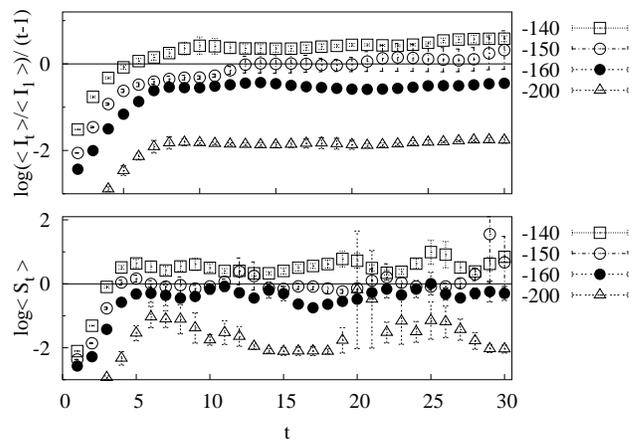}}
\caption{ Evolution of the median (top) and geometrical average
  (bottom) of the population of imaginary parts of the resolvent (see
  text) for $\lambda=-140$, $-150$, $-160$, and $-200$. Data for
  $N=10^6$.}
  \label{F-din}
\end{figure}

\begin{figure}
\centerline{ \includegraphics[angle =-90, width=\columnwidth]{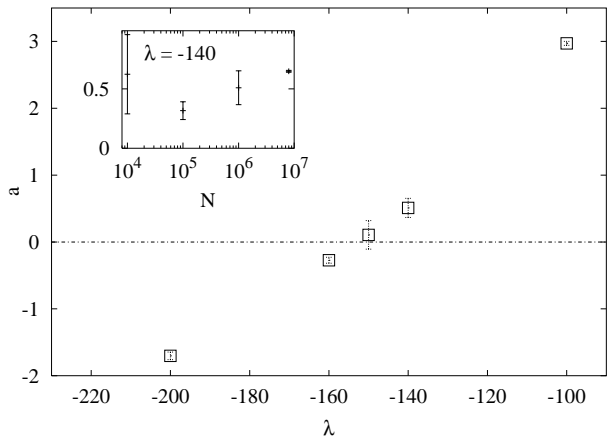} }
\caption{Values of the slope from a linear fit of $\langle \log I_t
  \rangle>$ for $t>10$. The mobility edge is estimated at
  $\lambda=-152(5)$. Inset: $N$-dependence of the slope.}
\label{F-edge}
\end{figure}

In search of a more reliable way to determine the threshold, we have
tried a different method, based on the observation that in the
localized region and in the $\varepsilon\to 0$ limit, $\text{Im}\,
{\bf R}_{ii}=0$ except for a vanishing fraction of particles,
suggesting that a population of real resolvents is dynamically stable
under Eq.~\ref{DINAMICA}. The idea \cite{Abouchacra} is to start with
an equilibrated population of real resolvents (setting $z=\lambda$),
add a small imaginary part and evolve the population with
eq.~\ref{DINAMICA} to see whether the imaginary parts grow (extended
phase) or tend to disappear (localized phase). We have analyzed in
this way the negative mobility edge at $\Gamma=1.1$: We obtained an
equilibrated population of real resolvents, then added an
infinitesimal imaginary part, which evolved with the linearized (with
respect to the imaginay part) form of Eq.~\ref{DINAMICA}. We evolved
the real part with Eq.~\ref{DINAMICA} as if the population were real.
Since the pdf of $\text{Im}\, {\bf R}_{jj}^{(t)}$ is extremely broad,
one must be careful in the quantity one chooses to examine. In
ref.~\onlinecite{Abouchacra} it was proposed to consider the quotient
$S_t$ of the geometric mean of the imaginary parts at times $t$ and
$t-1$. $S_t$ was assumed to be $t$-independent, and it was expected to
be equal to 1 at the mobility edge. However, the threshold found with
this method was far from theoretical estimates. Indeed, as can be seen
in Fig.~\ref{F-din}, this quantity has extremely large fluctuations,
that make it practically useless to determine the threshold. Instead,
we have found that the {\em median} (let us call it $I_t$) of the pdf
of $\text{Im}\, {\bf R}_{ii}^{(t)}$ is much more reliable. However,
obtaining the thermodynamic limit for $I_t$ requires $N$ that grows
fastly with $t$. For $t<15$ we have found no differences between
$N=10^6$ and $N=8\times 10^6$.  One would expect that $\log I_t \sim a
t$, with $a$ less than (greater than) zero in the localized (extended)
phase, and this is indeed what we find at large $t$
(Fig.~\ref{F-din}). For $N=10^6$ we have performed 10 trajectories,
then estimated the errors with a jacknife method.  A linear fit of
$\log I_t$, for $1\le t\le 15$ gives values $a$ that locate the
mobility edge at $\lambda=-152(5)$ (Fig.~\ref{F-edge}).

In summary, we have addresed the problem of Anderson localization in
Euclidean Random Matrices, specializing to a liquid Instantaneous
Normal Modes calculation. Under the random-graph approximation for
particle correlations, which keeps only the information of the radial
distribution function, a closed equation was found for the pdf of the
diagonal element of the resolvent matrix ${\bf R}_{ii}$. This equation
was solved numerically by means of a population dynamics algorithm. We
have shown that the random-graph approximation works rather well at
low density, but significantly overestimates the weight of unstable
modes when the system is close to crystallization. We have shown how
to locate the mobility edge from the mean of the square of the
imaginary part of ${\bf R}_{ii}(z)$. We have studied numerically the
stability of a population of real resolvents, improving over the
numerical method of Abou-Chacra {\sl et al.}~\cite{Abouchacra}, and
obtaining an estimate in reasonable agreement with the value obtained
with complex resolvents.

We acknowledge partial support from MCyT (Spain), contracts
FPA2001-1813, FPA2000-0956 and BFM2003-08532-C03 and ANPCyT
(Argentina). S.C. was supported by the ECHP programme under contract
HPRN-CT-2002-00307, {\em DYGLAGEMEM}. T.S.G. is career scientist of
CONICET (Argentina). V.M.-M. is a {\em Ram\'on y Cajal} research
fellow. P.V. was supported by the European Comission through contract
MCFI-2002-01262.


\end{document}